\input{aipcheck.tex}

\documentclass[sort&compress]{aipproc}

\def\lum{erg~s$^{-1}$}

\layoutstyle{8x11double}

\SetInternalRegister\hbadness{8000} 

\begin{document}

\title 
      [New BeppoSAX-WFC results on superbursts]
      {New BeppoSAX-WFC results on superbursts}


\author{J.J.M. in 't Zand}{
  address={SRON National Institute for Space Research,
           Sorbonnelaan 2, 3584 CA Utrecht, the Netherlands},
  altaddress={Astronomical Institute, Utrecht University, P.O. Box 80000,
                3508 TA Utrecht, the Netherlands}
}

\iftrue
\author{R. Cornelisse}{
  address={Dept. of Physics and Astronomy, University of Southampton,
                Hampshire SO17 1BJ, U.K.}
}

\author{E. Kuulkers}{
  address={ESTEC/ESA, SCI-SDG, Keplerlaan 1, 2201 AZ Noordwijk,
         the Netherlands}
}

\author{F. Verbunt}{
  address={Astronomical Institute, Utrecht University, P.O. Box 80000,
                3508 TA Utrecht, the Netherlands}
}

\author{J. Heise}{
  address={SRON National Institute for Space Research,
           Sorbonnelaan 2, 3584 CA Utrecht, the Netherlands},
  altaddress={Astronomical Institute, Utrecht University, P.O. Box 80000,
                3508 TA Utrecht, the Netherlands}
}
\fi

\copyrightyear  {2001}

\begin{abstract}
Presently seven superbursters have been identified representing 10\%
of the total Galactic X-ray burster population. Four superbursters
were discovered with the Wide Field Cameras (WFCs) on BeppoSAX and
three with the All-Sky Monitor and Proportional Counter Array on
RXTE. We discuss the properties of superbursters as derived from WFC
observations. There are two interesting conclusions. First, the {\em
average} recurrence time of superbursts among X-ray bursters that are
more luminous than 10\% of the Eddington limit is 1.5 yr per object.
Second, superbursters systematically have higher $\alpha$ values and
shorter ordinary bursts than most bursters that have not exhibited
superbursts, indicating a higher level of stable thermonuclear helium
burning. Theory predicts hitherto undetected superbursts from the most
luminous neutron stars.  We investigate the prospects for finding
these in GX~17+2.
\end{abstract}

\date{\today}

\maketitle

\section{Introduction}

\begin{figure}[t]
\caption{2-28 keV WFC light curves of the most recently published superburst
which came from 4U~1254-69. The top panel shows the light curve during
a 5-d long observation (gaps are earth occultations; low points
indicate dipping activity; 300-s resolution), the middle panel zooms
in at 8-s resolution, and the bottom panel at 2-s resolution.
\label{fig1254}}
\includegraphics[width=0.99\columnwidth,height=12cm]{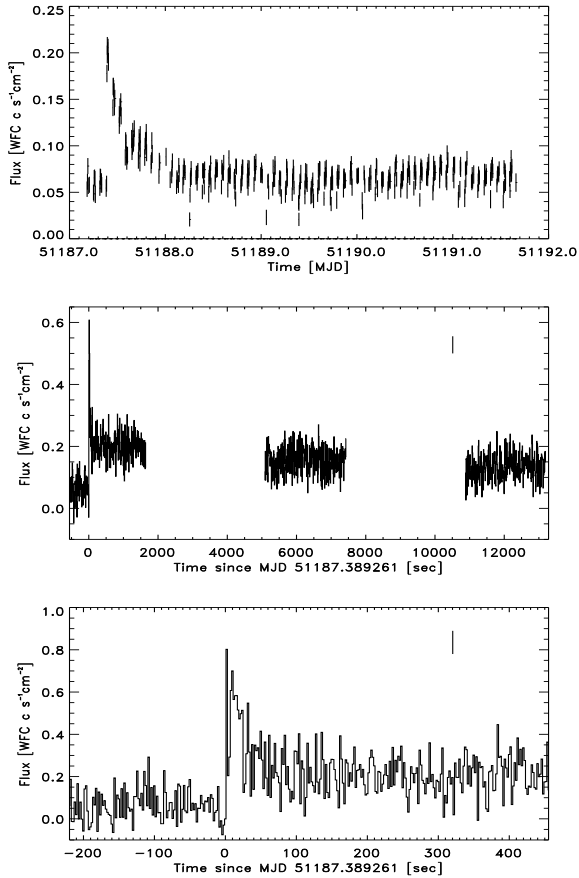}
\end{figure}

\begin{figure}[t]
\caption{2-28 keV light curves of the 4 WFC-detected superbursts so far,
in order of discovery. The dashed lines indicate the average pre-burst
flux level.
\label{figlc}}
\includegraphics[width=0.99\columnwidth]{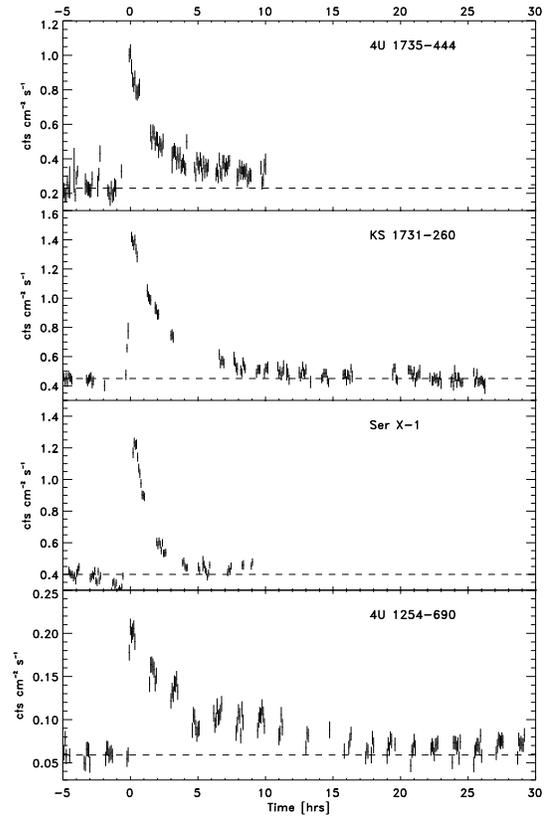}
\end{figure}

Three years ago the discovery of a new type of thermonuclear runaway
process on a neutron star was reported: that of unstable carbon
burning. The BeppoSAX Wide Field Cameras (WFCs) in 1996 detected a
flare from the X-ray burster 4U~1735-44 which, according to Cornelisse
et al. \cite{cor00}, was reminiscent of type-I X-ray bursts (fast
rise, exponential decay, black body spectrum, cooling during decay)
but was roughly 10$^3$ as long and energetic. A similar flare was
detected from 4U~1820-303 in 1999 with the Proportional Counter Array
on RXTE and Strohmayer
\& Brown \cite{str02a} proposed that the longevity was due to the fact
that not hydrogen or helium was being burned, like in ordinary bursts,
but carbon, thus following an early model for $\gamma-$ray bursts by
Woosley \& Taam \cite{woo76}. This proposal was further developed by a
number of authors, motivated also by the detection of five more superbursts
\cite{cor02,kuu02a,kuu02b,wij01,str02b}. Cumming \& Bildsten \cite{cum01}
introduced the ingredient of a heavy element ocean which relaxes the
carbon reservoir constraints and allows the recurrence time to be
substantially smaller than a decade. Recently, an eighth superburst was
discovered in archival WFC data (see Fig.~\ref{fig1254}) which
significantly increases the parameter range of superbursts
\cite{zan03a}. Superbursts have e-folding decay times between 1 and 6
hours, peak luminosities (after subtraction of the persistent
emission) between 0.4 and 3.4$\times10^{38}$~\lum, and occur on
neutron stars that accrete between 0.1 and 0.25 times Eddington (see
recent review by Kuulkers \cite{kuu03}; for a compilation of the
4 WFC-detected superbursts see Fig.~\ref{figlc}).

As pointed out by Strohmayer \cite{str02c}, superbursts provide the
prospect of powerful diagnostics of neutron stars in low-mass X-ray
binaries (LMXBs).  A quick TOO turn around could provide enhanced
statistics to find and accurately measure narrow spectral features and
determine from their gravitational redshift constraints on the neutron
star mass-radius relation (like was done with XMM-Newton observations
of EXO~0748-676 \cite{cot02}). Also, they may reveal constraints on
the binary orbit through Doppler shifts of the burst oscillation
frequency. This was partly demonstrated in 4U~1636-536
\cite{str02b}.

Despite its demise in May 2002, (archival) data from the BeppoSAX Wide
Field Cameras are still actively pursued. For a small part that is
because for a few percent of the observations the raw data processing
was only recently accomplished. Another motivation is that more
complex data are now being tackled with more sophisticated soft and
hardware. In light of this, new results have been obtained with
respect to superbursts which are discussed here (see also
\cite{zan03a}).

\section{Recurrence time}

The number of superbursts is now up to a level that the recurrence
time can be better constrained. This parameter is important for
fine-tuning the superburst model. It may particularly provide better
constraints on the mass fraction of the carbon in the flash layer and
the fraction of the liberated energy being carried away by neutrinos
(e.g., \cite{str02a,cum01}). Thus far recurrence has been observed in
only one source: 4U~1636-536. Initially, an interval time of 4.7 yr
was observed \cite{wij01}, but now a third superburst has been found
between the previous two resulting in recurrence times of 1.8
and 2.9 yr \cite{kuu04}. The question is whether these values are
representative for all superbursters.

Currently, seventy-six X-ray bursters are known in our Galaxy
\cite{zan03b}. Twenty-seven of these (give or take two) have been
persistent sources for at least 10 years (we include the long-duration
transients KS 1731-260 and 4U 1724-307). The seven superbursters are
part of this group.  The WFCs have extensive coverage of these
objects. The net exposure times summed over all observations and
sources is 7.9 yr (i.e., the average time per object is 0.3 yr). If
all 27 objects are identical in their superburst behavior, the implied
superburst recurrence rate is once per 2 yr. However, theory predicts
\cite{cum01} that sources with luminosities below 0.1 times Eddington will
not exhibit superbursts. We estimate the number of remaining bursters
at 18.  The WFC exposure on these is 5.9 yr, implying a superburst
recurrence time of 1.5 yr. If we go one step further and exclude those
systems that have luminosities larger than 0.25 times Eddington (the
Z-sources GX 17+2 and Cyg X-2, and Cir X-1), for which presumably
different recurrence times apply \cite{cum01}, the average recurrence
time is 1.2 yr. These recurrence times are smaller than expected
(e.g., \cite{cum01} and \cite{cum03a}).

\section{$\alpha$ and stable helium burning}

In order to produce sufficient amounts of carbon to fuel a superburst,
one needs to burn helium for a sufficiently long time and avoid that
the carbon is destroyed by subsequent proton and alpha captures, and
breakout reactions from the hot CNO cycle \cite{wal81,woo03,cum03a}.
The manners in which carbon may be destroyed imply that preferably 1)
the hydrogen abundance should be at a minimum level in the burning
mixture and 2) the helium should be burned in a stable manner to avoid
the temperature to rise above the threshold for initiation of the hot
CNO cycle as happens in unstable burning
\cite{woo03} (in other words, stable helium burning would produce 
the carbon while (un)stable hydrogen burning in another layer would
produce the heavy elements that provide ignition conditions for
smaller amounts of carbon as proposed in \cite{cum01}). Diagnostics
for the hydrogen content in the helium-burning layer and stability of
the burning are provided by the so-called $\alpha$ parameter and the
duration of ordinary bursts.

$\alpha$ is defined as the ratio of the integrated radiation energy
between two consecutive bursts and the fluence of the burst concluding
this interval.  The persistent emission is due to the release of
gravitational energy in the accretion disk around the neutron
star. Per nucleon this amount to $\approx$200 MeV. The energy released
in thermonuclear burning depends on the chain of nuclear reactions.
When hydrogen is burned, the rapid proton capture process dominates
the energy production at 7 MeV per nucleon. When hydrogen is absent,
the triple alpha process dominates the energy production at 1.6 MeV
per nucleon. If the gravitational and thermonuclear energy production
transforms solely to isotropic radiation, $\alpha$ should be about 30
for hydrogen-dominated bursts and 130 for helium-dominated bursts.

Accurate $\alpha$ determinations have been made for about 10 systems
with EXOSAT \cite{jvp88a}. Van Paradijs et al. \cite{jvp88b} noticed
an interesting measurement that $\alpha$ occasionally is very large in
4U~1735-44, namely almost 8000. They attributed this in part to stable
helium burning, despite the fact that this is inconsistent with theory
\cite{fuj81}.

The accurate measurement of $\alpha$ is difficult. Firstly, continuous
coverage is needed between bursts and this is difficult for
observations from low-earth orbit satellites given that typical burst
recurrence times are hours. This explains the success of the high
flying EXOSAT. Secondly, one needs broad-band spectral coverage to
measure with a reasonable accuracy the bolometric flux.  Particularly
for the persistent emission this is relevant, since often a
considerable fraction of the flux is outside the typical 2-10~keV
bandpass. Thirdly, one needs sufficient sensitivity to be able to also
detect the weak X-ray bursts. For LMXBs that are at distances smaller
than the canonical 8~kpc this is not really an issue, even for
modest-sized instruments, but for distances beyond that smaller
instruments become insufficient. These issues can only be resolved if
one uses an instrument in a high-earth orbit with coverage up to
100~keV in a staring observation of at least a day duration. Possibly
INTEGRAL observations could contribute to accurate $\alpha$
measurements.

\begin{table}[t]

\begin{tabular}{lrr}
\hline\hline
Object name  & $\alpha^{\rm (a)}$ & $\tau^{\rm (b)}$ [sec]\\
\hline	     
4U~1254-690  & 4800   & $6.0\pm2.0$       \\
4U~1636-536  &  440   & $6.2\pm0.1$   \\
KS~1731-260$^{\rm (c)}$  &780     & $5.6\pm0.2$   \\
4U~1735-444  & 4400   & $3.2\pm0.3$   \\
GX~3+1       & 2100   & $4.6\pm0.1$   \\
4U~1820-303  & 2200   & $4.5\pm0.2$   \\
Ser X-1      & 5800   & $5.7\pm0.9$    \\
\hline              
EXO~0748-676 &  140   & $12.8\pm0.4$ \\
4U~1702-429  &   58   & $7.7\pm0.2$  \\
4U~1705-44   & 1600   & $8.7\pm0.4$   \\
GX~354--0    &   97   & $4.7\pm0.1$  \\
A~1742-294   &  130   & $16.8\pm1.0$ \\
GS~1826-24   &   32   & $30.8\pm1.5$ \\
\hline\hline
\multicolumn{3}{l}{
\parbox[t]{0.95\columnwidth}{
$^{\rm (a)}\alpha$ is ratio of average persistent 2--28 keV flux (in
WFC c~s$^{-1}$cm$^{-2}$) times average wait time between two bursts
(2nd column) and burst fluence (in WFC c~cm$^{-2}$); $^{\rm
(b)}$e-folding decay time of the average 2--28 keV burst profile;
$^{\rm (c)}$This is a transient and only data are given for persistent
flux levels comparable to when the superburst occurred.}}
\end{tabular}

\caption{Average burst properties of all superbursters
 (above the dividing line) and six non-superbursters, as observed with
BeppoSAX-WFC. From \cite{zan03a}.\label{tab2}}

\end{table}

Alternatively, one may resort to statistical studies. If large numbers
of bursts are detected while a system is in the same bursting regime,
the average bursting rate in combination with the average persistent
and burst spectrum will provide a reasonable estimate of
$\alpha$. This kind of data is abundantly provided by the WFCs for the
majority of X-ray bursters (e.g., \cite{cor03} and \cite{zan03b}).  A
systematic spectral analysis of all persistent and burst data has not
been performed yet, but a simplified definition of $\alpha$ alleviates
this shortcoming. Instead of the energy fluence ratio, we use the
observed photon fluence ratio. The usefulness of this definition has
been confirmed in \cite{zan03a} and, in fact, the photon-based
$\alpha$ values do not differ by more than a few tens of percents of
the energy-based values. There is a small caveat: the WFCs are only
sensitive enough to peak fluxes roughly brighter than 0.3 Crab. This
may artificially overestimate $\alpha$ somewhat in 4U~1254-690.  We
list in table~\ref{tab2} the results, separated for confirmed
superbursters and others. We also list the e-folding decay times for
the average burst profiles. This decay time is a good diagnostic for
the relative amount of hydrogen burning in the burst. With one
exception, the table shows a clear dichotomy between superbursters and
other bursters: for superbursters $\alpha$ is high and the decay time
short. This strongly suggests that unstable helium burning occurs in a
hydrogen-poor environment and that stable helium burning is
important. This for the first time provides observational evidence for
a clear difference in non-superburst characteristics between
superbursters and other bursters.

The one exception, 4U~1705-44, should provide a test to the predictive
power of this diagnostic. So far, no superburst has been observed from
this system with the WFCs nor RXTE/ASM.

\section{Testing a prediction: superbursts in high-luminosity systems}

\begin{figure}[t]
\caption{2-28 keV light curve and time-resolved spectroscopy of a
likely first superburst from GX 17+2. Before the spectroscopy, the
persistent spectrum was subtracted as determined from before these
data.\label{figgx}}
\includegraphics[width=0.99\columnwidth]{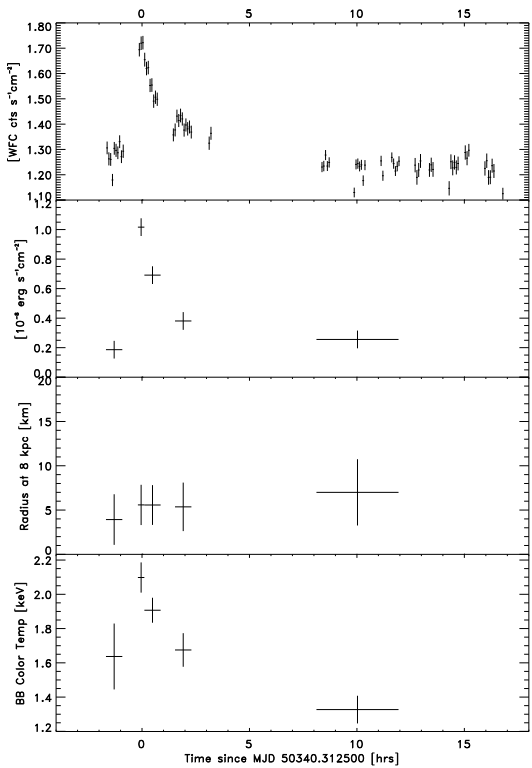}
\end{figure}

Cumming \& Bildsten \cite{cum01} predict that for a carbon mass
fraction $X_{12}<0.1$ any neutron star accreting faster than 0.1 times
the Eddington limit should exhibit superbursts. For a mass accretion
rate close to Eddington, the recurrence time is predicted to be a few
weeks and the cooling time scale about 1~hr.  The seven superbursters
discovered up to now accrete at most at one quarter of the Eddington
limit. As noted by Cumming \& Bildsten \cite{cum01}, this may be a
selection effect. The dynamic range available for superbursts in
near-Eddington systems is much less than in weaker systems: if the
system is emitting at 90\% of Eddington, the signal-to-background
ratio is 0.1 at maximum while it may be 10 if the persistent
luminosity level is near 10\% of Eddington. Furthermore, the amplitude
of the variability in the more luminous systems may be similar to the
peak flux of superbursts.

In order to test the theory, we initiated a search in WFC data for
superbursts in one of the few well-known persistently high-luminosity
X-ray bursters, GX 17+2. It appears that indeed this system exhibits
superbursts. An example is presented in Fig.~\ref{figgx}.  The
e-folding decay time of this flare is 1.9 hr, which is a factor of 25
longer than the longest ordinary burst observed from GX 17+2 thus far
(which is already relatively long for an ordinary burst with a decay
time of 4.5 minutes;
\cite{kuu02c}). The spectrum is consistent with black body emission
showing cooling during the decay. The fast rise of the flare candidate
seems markedly different from the flares commonly observed in this
Z-source. Thus, this appears a genuine superburst which would fit
theoretical predictions excellently. We are continuing to analyze the
WFC data in detail to find more superburst candidates from this source
and fine tune discriminating diagnostics against accretion-type
flares.  Furthermore, we are pursuing data from other persistently
high-luminosity LMXB bursters such as Cyg X-2.

\begin{theacknowledgments}

We thank Andrew Cumming for useful discussions. This work is financially
supported by the Netherlands Organization for Scientific Research (NWO).

\end{theacknowledgments}


\end{document}